\documentclass[12pt]{iopart}
\usepackage{graphicx}
\begin{document}
\letter{On stability of the neutron rich Oxygen isotopes}

\author{K.A.~Gridnev$^{1,2}$,\ D.K.~Gridnev$^{1,2}$,\
V.G.~Kartavenko$^{2,3}$\footnote[3]{To whom correspondence should
be addressed(kart@thsun1.jinr.ru)},\ V.E.~Mitroshin,\
V.N.~Tarasov$^{5}$,\ D.V.~Tarasov$^{5}$\ and W.~Greiner$^{2}$}

\address{\dag\ Institute of Physics, St.~Petersburg State University }

\address{$^{2}$\ Frankfurt Advance Study Institute,
J.W.G. University, Frankfurt, Germany}

\address{$^{3}$\ Joint Institute for Nuclear Research, Dubna,
Russia}

\address{$^{4}$\ Kharkov National Univeristy, Ukraine}

\address{$^{5}$\ Kharkov National Scientific Center HFTI, Ukraine}

\begin{abstract}
Stability with respect to neutron emission is studied for highly
neutron-excessive Oxygen isotopes in the framework of
Hartree-Fock-Bogoliubov approach with Skyrme forces Sly4 and Ska.
Our calculations show increase of stability around $^{40}$O.

\end{abstract}



\submitto{\JPG}

\maketitle

One of the major challenges of nuclear physics is to enlarge the
present limits of the chart of the nuclides~\cite{kemer03,
nsac2002}. The experimental progress during last decade has
allowed to reach the proton drip-line up to charge
80~\cite{exon04, mittig04}. In contrast, the situation is quite
different on the neutron-rich side. Recent studies of neutron-rich
light nuclei have shown very exciting issues, that new magic
numbers might appear and some disappear when moving to the neutron
drip-line~\cite{jonson04}. Near the neutron drip-line, the
neutron--matter distribution becomes very diffuse and of large
size giving rise to "neutron halos" and "neutron skins". These
neutron-rich objects have sparked renewed interest in the nuclear
structure theory~\cite{vretenar04, tou03}.

In this Letter we present our preliminary results in searching for
highly neutron-excessive stable light nuclei within
Hartree-Fock-Bogoliubov (HFB) approach with Skyrme effective
interaction, which has the following form
\begin{eqnarray}\label{skyr}
    V_{ij}  = t_0 (1 + x_0 P_{\sigma}) \delta({\bf r}) + (1/2)
    t_1 (1 + x_1 P_{\sigma}) [{\bf k'}^2 \delta({\bf r}) + \nonumber \\
    \delta({\bf r}) {\bf k}^2
    ] +
    t_2 (1+ x_2 P_{\sigma}){\bf k'} \delta({\bf r}){\bf k} +
    (1/6) t_3 (1 +
    x_3 P_{\sigma}) \nonumber\\
    \rho^{\alpha} ({\bf R}) \delta({\bf
    r}) + i W_0 [{\bf k'} \times \delta({\bf r}){\bf k}] (\sigma_i + \sigma_j
    ) \nonumber
    \end{eqnarray}
where ${\bf r} = {\bf r}_i - {\bf r}_j$, ${\bf R} = ({\bf r}_i +
{\bf r}_j )/2$ , ${\bf k} = - i (\overrightarrow{\nabla_i} -
\overrightarrow{\nabla_j })/2$, ${\bf k'} =  i
(\overleftarrow{\nabla_i} - \overleftarrow{\nabla_j })/2$,
$P_\sigma = (1 + \sigma_i \sigma_j )/2$. Parameters are given in
Table~1.
%

\hspace*{-10mm}\begin{table}[h] \caption{\label{tab1}Parameters of
the Skyrme forces}
%
\begin{tabular}{@{}*{11}{c}} \br {\small Force} &
$t_0$ & $t_1$ & $t_2$ & $t_3$ & $x_0$ & $x_1$ & $x_2$ & $x_3$ &
$W_0$ & $\alpha$ \cr \br
    &{\tiny MeV fm$^3$} &{\tiny MeV fm$^5$} &{\tiny MeV fm$^5$} &{\tiny MeV fm$^{3+3\alpha}$} &   &   &   &   & {\tiny MeV fm$^5$} &
  \cr \br
  Sly4 & -2488.9 & 486.82 & -546.39 & 13777.0 & 0.834 & -0.344 & -1.0 & 1.354 & 123.0 &1/6
  \cr
  Ska & -1602.8 & 570.9 & -67.70 & 8000.0 & -.020 & 0.0 & 0.0 & -0.286 & 125.0 &1/3
  \cr \br
\end{tabular}

\end{table}
We have used the set of parameters Ska~\cite{Ska} and compared the
results with the most widely used set Sly4~\cite{SLy}. In
\cite{Gontchar} some of us shown that for deformed nuclei
$^{25}$Mg and $^{29-31}$Si the most satisfactory description of
observed spectra comes with the set Ska. Pairing effects were
included in the standard way with the pairing constant $G = 19/A$
both for protons and neutrons and restricted to the space of
bounded one-particle states.
\begin{figure}[htbp]
\includegraphics[width=7.5cm]{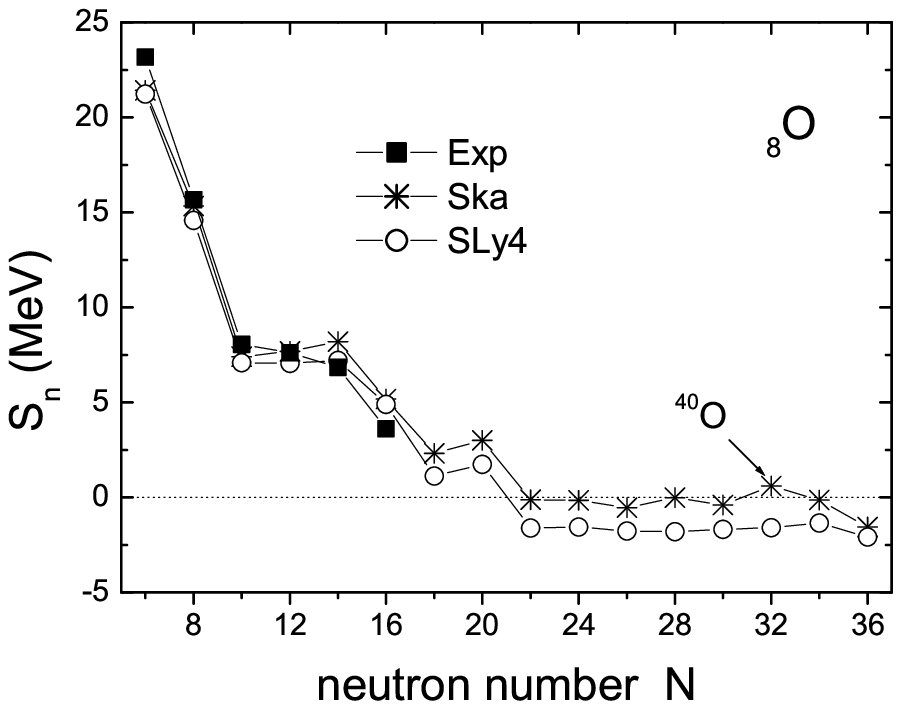}
\includegraphics[width=7.5cm]{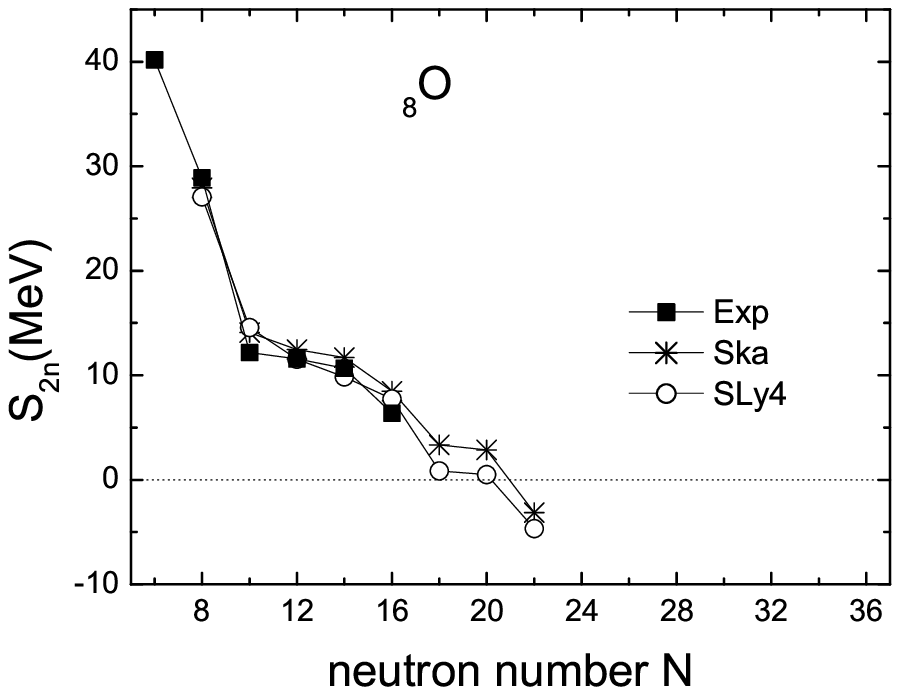}
\caption{Calculated separation energies of one ($S_n$) and two
($S_{2n}$) neutrons for isotopes $^{14-44}$O with different
choices of Skyrme forces compared to the experimental
data~\cite{expmass}. We did not plot the two-neutron separation
energy for $^{40}$O because all the neighboring isotopes are
unstable.}
\label{fig:1}       
\end{figure}

We carried out the investigations of isotopes $^{4-12}$He,
$^{14-44}$O and $^{38-80}$Ca and analyzed how results depend on
forces we have used. For isotopes $^{4-12}$He our results matched
the known ones \cite{sdnpd03}. For Helium the last stable isotope
with respect to two-neutron emission is $^8$He. Comparison with
experiment (see the figures) shows that both Ska and Sly4 equally
good describe the known experimental data. In our calculations
with forces Ska we have found a nucleon stable isotope $^{40}$O
(See Fig.~\ref{fig:1}). We did not plot the two-neutron separation
energy for $^{40}$O because all its neighboring isotopes are
unstable. With forces Sly4 this isotope appears to be unstable
with respect to one neutron emission, though the last filled level
is close to zero and one can talk about ``quasistability'' in this
case. From nucleus to nucleus the situation repeats itself,
whenever the nucleus is ``quasistable'' with interactions Ska,
then it is ``quasistable'' with interactions Sly4. In all
investigated cases the last filled level for nuclei close to
nucleon stability borderline always had a negative parity. And
under ``quasistable'' we mean that this nucleon has a non-zero
orbital momentum and the resulting centrifugal barrier prevents
the neutron from emission at its low energies.

The obtained data for $^{40}$O is given in Table~2. One can see
that the values of proton and neutron deformation for $^{40}$O are
negligibly small.
\begin{table}[h]
\caption{ \label{tab:2}Calculated values of binding energy $E$,
neutron and proton separation energy $S_{p,n}$, root mean square
radii $r_{p,n}$ quadrupole moments $Q_{p,n}$ and deformation
parameters $\beta_{2}^{p,n}$ for the stable isotope $^{40}$O as
calculated with Ska forces.}
\begin{indented}
\lineup \item[]\begin{tabular}{@{}*{9}{c}} \br $E$ & $S_n$ & $S_p$
& $r_n$ & $r_p$ & $Q_n$ & $Q_p$ & $\beta_{2}^n$ & $\beta_{2}^p$
\cr \mr
  MeV & MeV & MeV & fm & fm & e$^2$ fm$^2$ & e$^2$ fm$^2$ &  &  \cr
  168.274 & 0.593 & 36.822 & 4.202 & 2.943 & 0.031 & 0.003 & 0.004 & 0.004
  \cr \br
\end{tabular}
\end{indented}
\end{table}

\begin{figure}[htbp]
\begin{center}
\includegraphics[width=10cm]{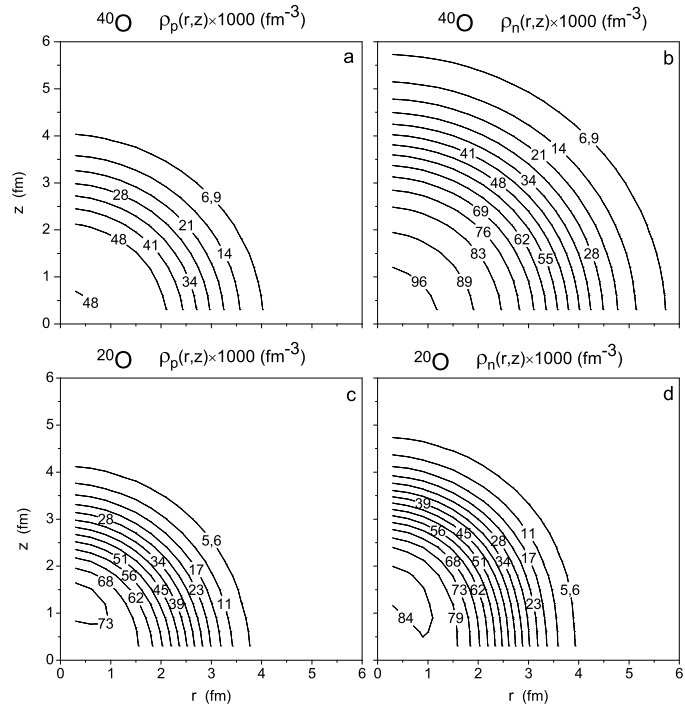}
\end{center}
\caption{The map of proton $\rho_p$ and neutron $\rho_n$
distributions calculated for $^{40}$O (a,b) and for $^{20}$O
(c,d). The proton ``cloud'' is expanding with the increasing of
the neutron number, but it remains coated with the neutron halo
with 2 fm thickness.} \label{fig:density}
\end{figure}

The maps of proton and neutron distributions in $r,z$ coordinates
(incorporating the symmetry of the problem) for $^{40}$O and
$^{20}$O are shown in Fig.~\ref{fig:density}.  One can see
although the proton ``cloud'' is expanding it remains coated with
the neutron halo which is about 2 fm thick.

\begin{figure}[htbp]
\begin{center}
\includegraphics[width=9cm]{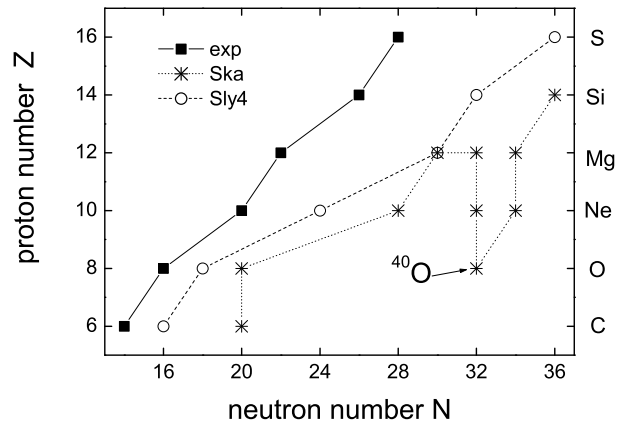}
\end{center}
\caption{The part of the neutron drip-line. For each value of Z
the heaviest stable isotope has been presented for experimental
data (Audi 2003~\cite{expmass}) and Sly4 forces.\\
For the very neutron-rich isotopes HFB calculations with Ska
forces shows the different behaviour of the neutron drip-line. See
Fig.~\ref{fig:1} for the details.} \label{fig:limits}
\end{figure}

Our preliminary calculations of nearby isotopes showed, that there
are stable isotopes around $^{40}$O among even-even nuclei, namely
$^{40, 42, 44}$Ne with one neutron separation energies are
respectively $S_n = 0.13, 0.43, 0.1$ MeV and $^{44, 46}$Mg with
one neutron separation energies $S_{n} = 0.8, 0.67$ MeV (See
Figure~\ref{fig:limits} for the details). These stable isotopes
were also found with Ska forces and except $^{44}$Mg they lie
beyond the conventional stability valley. We have observed that
the usual shell closure at N=50 is practically absent for the
neutron rich Ca isotopes, and at N=40 a new closure appears.
%

It should be mentioned, that the stability with respect to neutron
emission is defined within very narrow range of the binding energy
(about 0.5~MeV). Therefore it would be desirable to check this
effect with other theoretical models.

Our detailed investigations of the decay properties of the neutron
drip-line nuclei in this region of possible stability (isotopes of
C,O,N,Mg with the very large neutron excess) are in progress. The
results will be published elsewhere.

\subsection*{Acknowledgements}
{T}his work was partially supported
by \uppercase{D}eutsche \uppercase{F}orschungsgemeinschaft
   (grant 436 \uppercase{RUS} 113/24/0-4), \uppercase{R}ussian \uppercase{F}oundation
   for \uppercase{B}asic \uppercase{R}esearch
   (grant 03-02-04021) and the \uppercase{H}eisenberg-\uppercase{L}andau
   \uppercase{P}rogram (\uppercase{JINR, D}ubna). D.K. Gridnev appreciates the financial support from the
Humboldt Foundation.

\subsection*{References}
\label{refs}



\begin{thebibliography}{}
%
%
\bibitem{kemer03}See e.g. {\em Procs. NATO ADV. STUDY INST., Structure
and Dynamics of Elementary Matter}, 22 September –- 2 October
2003, Kemer, Turkey. Eds. W.~Greiner, M.G.~Itkis, J.~Reinhardt and
M.C.~G\"ucl\"u. Nato Science Series. II. Math., Phys. and
Chemistry.   Kluwer Academic Publishers, London. Vol.~166, 2004.
%
\bibitem{nsac2002} Opportunities in Nuclear Science, A Long-Range Plan for
the Next Decade, DOE/NSF Nuclear Science Advisory Committee, April
2002, published by U.S. Dept. of Energy.\\
http://www.sc.doe.gov/henp/np/nsac/nsac.html
%
\bibitem{exon04}
International Symposium on Exotic Nuclei "EXON - 2004" Peterhof,
Lakes Onega and Ladoga, 5-12 July 2004. World Scientific in press
%
\bibitem{mittig04}W.~Mittig, P.~Roussel-Chomaz and A.C.C.~Villari, {\em
Europhysics News}, {\bf 35} (2004) No.4, p.1.
%
\bibitem{jonson04} B.~Jonson, {\em Phys. Peports}, {\bf 389} (2004)
pp.1--59.
%
\bibitem{vretenar04} {\em See e.g. review} D.~Vretenar,
arXiv:nucl-th/0409060 (2004).
%
\bibitem{tou03} E.~Teran, V.E.~Oberacker, and A.S.~Umar, Phys.
Rev. C, {\bf 67} (2003) 064314.
%
\bibitem{Ska}
H.S. Kohler, Nucl.~Phys. {\bf A258}, (1976) 301
%
\bibitem{SLy}
E. Chabanat, P~Bonche, P.~Haensel, J.~Meyer, and F.~Shaeffer,
Nucl.~Phys. {\bf A635}, (1998) 231; Nucl.~Phys. {\bf A643}, (1998)
441
%
\bibitem{Gontchar}
Yu.V.~Gontchar {\em et.al.}, Yad.~Fiz. {\bf 41} (1985) 590
%
\bibitem{sdnpd03}
%
M.V. Stoitsov, J.~Dobaczewski, W.~Nazarewicz, S.~Pittel,
   and D.J.~Dean,\\ Phys.~Rev.~C, {\bf 68} (2003) 054312.
%
\bibitem{expmass}G.~Audi, A.H.~Wapstra, and C.~Thibault,
{Nucl. Phys.} {A729} (2003) 337.
%
\end{thebibliography}
\end{document}